\documentstyle[12pt]{ets2000}
\def\rfrs#1#2{eqs.(\ref{#1})-(\ref{#2})}

\def\eqi{\begin{equation}}
\def\eqf{\end{equation}}
\def\eqia{\begin{eqnarray}}
\def\eqfa{\end{eqnarray}}
\def\lb#1{\label{#1}}

\begin{document}
\title{
Is it possible to test directly General Relativity in the
gravitational field of the Moon? }
\author{
Lorenzo Iorio$^{1)}$
\thanks{Fax: +390805443144, E-mail: Lorenzo.Iorio@ba.infn.it }}
\affil{
1) Dipartimento Interateneo di Fisica, \\
Via Amendola 173, 70126, Italy}
\footnotetext{ Fax: +390805443144, E-mail:
Lorenzo.Iorio@ba.infn.it }
\begin{abstract}
In this paper the possibility of measuring some general
relativistic effects in the gravitational field of the Moon via
selenodetic missions, with particular emphasis to the future
Japanese SELENE mission, is investigated. For a typical
selenodetic orbital configuration the post-Newtonian
Lense-Thirring gravitomagnetic and the Einstein's gravitoelectric
effects on the satellites orbits are calculated and compared to
the present-day orbit accuracy of lunar missions. It turns out
that for SELENE's Main Orbiter, at present, the gravitoelectric
periselenium shift, which is the largest general relativistic
effect, is 1 or 2 orders of magnitude smaller than the
experimental sensitivity. The systematic error induced by the
mismodelled classical periselenium precession due to the first
even zonal harmonic $J_2$ of the Moon's non-spherical
gravitational potential is 3 orders of magnitude larger than the
general relativistic gravitoelectric precession.
\end{abstract}
\section{Introduction}
Some of the most interesting effects that general relativity
predicts for the orbit of a test body in the field of a central
mass are the secular Lense-Thirring gravitomagnetic precessions of
the node $\Omega$ and the pericenter $\omega$, generated by the
off-diagonal terms of the stationary part of the spacetime metric
(Lense and Thirring, 1918; Ciufolini and Wheeler, 1995), and the
secular Einstein's gravitoelectric precession of the pericenter,
induced by the Schwarzschild's static part of the spacetime metric
(Ciufolini and Wheeler, 1995). Their expressions, in the
weak-field and slow-motion approximation, referred to an
asymptotically inertial frame of reference with its origin in the
center of mass of the central body,
are \eqia \dot\Omega_{\rm GM} & = & \frac{2GJ}{c^{2}a^{3}(1-e^{2})^{\frac{3}{2}}},\lb{letinodo}\\
\dot\omega_{\rm GM} & = & -\frac{6GJ\cos i}{c^{2}a^{3}(1-e^{2})^{\frac{3}{2}}},\lb{letiperi}\\
\dot\omega_{\rm GE} & = & \frac{3nGM}{c^{2}a(1-e^{2})},\lb{scva}
\eqfa in which $G$ is the Newtonian gravitational constant, $c$ is
the speed of light in vacuum, $J$ and $M$ are the proper angular
momentum and the mass, respectively, of the central body, $a$, $e$
and $i$ are the semimajor axis, the eccentricity and the
inclination, respectively, of the orbit of the test mass and
$n=\sqrt{\frac{GM}{a^{3}}}$ is its mean motion.

In the weak field of solar system such features are very tiny and
difficult to measure because of lots of other competing classical
effects which in many cases are quite larger than them. To-day,
the Lense-Thirring effect has been measured, with $20\%$ accuracy,
in the gravitational field of the Earth by means of the
laser-ranged data to LAGEOS and LAGEOS II satellites (Ciufolini,
2000). Moreover, the ambitious GP-B mission (Everitt et al.,
2001), which is scheduled to be launched in spring 2002, will
measure the effect of the gravitomagnetic field of the Earth on
orbiting gyroscopes at a claimed precision of 1$\%$. The
Einstein's precession of the pericenter has been detected in the
well-known radar ranging measurements of the Mercury's perihelion
advance in the field of the Sun with $1\%$ or better accuracy
(Shapiro, 1990).

In general, it should be pointed out that the experimental basis
of General Relativity is not yet particularly wide because, from
one hand, it is difficult to work out new predictions of the
theory which could be effectively detected, and, from the other
hand, it is not easy to perform experiments, even for those
effects which are already well known theoretically, or which have
been already tested in some particular contexts. The
gravitomagnetic Lense--Thirring drag and the gravitoelectric
Einstein pericenter shift fall in this category, so that, in the
author's opinion, every effort aimed to the investigation of the
possibility of enlarging the experimental foundations of General
Relativity in various contexts and by using different techniques
deserves attention.

Recent years have seen increasing efforts in selenodetic activity
conducted by several lunar missions like Clementine and Lunar
Prospector. One of the main goals of such missions is the mapping
of the lunar gravity field (Lemoine et al., 1997; Konopoliv et
al., 1998; 2001). Such kinds of selenodetic missions seem to be
just the natural candidates in order to detect direct general
relativistic effects of the lunar gravitational field on the orbit
of test bodies. Indeed, about the Moon dynamics, we can speak, in
a global sense, of the motions of the Moon's center of mass and,
in a local sense, of the motions around its center of mass.
Recently, many general relativistic features of the global motion
of the Moon in the Sun's gravitational field, among which the most
famous is the geodetic or de Sitter precession, have been
investigated both theoretically (Mashhoon and Theiss, 1991; 2001)
and experimentally (Ciufolini, 2001; Nordvedt, 2001). In this
paper we investigate quantitatively the possibility of measuring
the impact of General Relativity in the local dynamics of a
satellite--Moon system with particular emphasis to the SELENE
mission. Knowing the present--day level of experimental
sensitivity to two of the most important general relativistic
direct effects in the Moon's gravitational field may help to focus
further attentions to the lunar arena or to redirect them to
other, more feasible and promising scenarios.

The plan of the work is the following. In section 2 the
relativistic effects on the SELENE subsatellites are analyzed.  In
section 3 their values are compared to the obtainable experimental
accuracy. Section 4 is devoted to the conclusions.

\section{Relativistic effects on the orbits of the SELENE satellites}
The forthcoming Japanese SELENE mission (Ohta et al., 2000) is of
relevant scientific interest. It is currently under development by
National Space Development Agency of Japan (NASDA) and Institute
of Space and Astronautical Science (ISAS), and will be launched in
2004 to elucidate lunar origin and evolution. SELENE is composed
of Main Orbiter, and two micro sub-satellites; the Relay Satellite
(Rstar) and the VLBI Radio Satellite (Vstar) which will be used
for selenodesy experiments. The range-rate accuracy for the three
subsatellites should amount to $10^{-1}$ mm s$^{-1}$ (Heki et al.,
1999) (sampling interval 20-60 seconds, Doppler). The designed
nominal orbital parameters for Main Orbiter and Rstar are in
Tab.1. Their lifetimes should amount to 1-3 years.
\begin{table}[h]
\caption{Orbital parameters of SELENE sub-satellites.}
\begin{center}
\begin{tabular}{lllll}
\noalign{\hrule height 1.5pt} Orbital Parameter & Main Orbiter &
Rstar & Units\\ \hline \textit{a} & $1838$ & $3000$ & km \\
\textit{e} & 0.01 & 0.38 & -\\
\textit{i} & 95 & 95 & deg \\
\textit{n} & $8.83\times 10^{-4}$ & $4.26\times
10^{-4}$ & s$^{-1}$ \\
\noalign{\hrule height 1.5pt}
\end{tabular}
\end{center}
\end{table}
By using in \rfrs{letinodo}{scva} the values of Tab. 1 and those
for the Moon quoted in Tab. 2  it is possible to obtain for the
relativistic effects on the SELENE sub-satellites' orbits the
values quoted in Tab. 3. They hold in a selenocentric inertial
frame.
\begin{table}[h]
\caption{Moon parameters.}
\begin{center}
\begin{tabular}{llll}
\noalign{\hrule height 1.5pt} Parameter & Value &
Units \\ \hline \textit{M} mass& $7.349\times 10^{25}$ & g \\
\textit{J} proper angular momentum& $2.32\times 10^{36}$ & g cm$^{2} s^{-1}$\\
\textit{GM} & 4.9$\times 10^{18}$ & cm$^{3}$ s$^{-2}$ \\
\textit{R} radius& $1.738\times 10^{8}$ & cm\\
$\alpha$ proper angular velocity& 2.66$\times 10^{-6}$ & rad
s$^{-1}$\\
\textit{I} moment of inertia& 8.74$\times 10^{41}$ & g cm$^{2}$\\
$\frac{GM}{c^{2}}$ & $5.452\times 10^{-3}$ & cm\\
$\frac{GJ}{c^2}$ & $1.71\times 10^{8}$ & cm$^{3}$ s$^{-1}$\\
{\it $J_2$} mass quadrupole moment & $2.03428\times 10^{-4}$ & -\\
{\it $\delta J_2$} & $9\times 10^{-8}$ & -\\
 \noalign{\hrule
height 1.5pt}
\end{tabular}
\end{center}
\end{table}
\begin{table}[h]
\caption{Relativistic effects on the nodes and periselenia of
SELENE sub-satellites.}
\begin{center}
\begin{tabular}{lllll}
\noalign{\hrule height 1.5pt} Relativistic precession & Main
Orbiter &
Rstar & Units\\ \hline $\dot\Omega_{\rm GM}$ & $5.4\times 10^{-17}$ & $1.2\times 10^{-17}$ & rad s$^{-1}$ \\
$\dot\omega_{\rm GM}$ & $1.4\times 10^{-17}$ & $3.3\times 10^{-18}$ & rad s$^{-1}$ \\
$\dot\omega_{\rm GE}$  & $7.8\times 10^{-14}$ & $3.1\times 10^{-14}$ & rad s$^{-1}$ \\
\noalign{\hrule height 1.5pt}
\end{tabular}
\end{center}
\end{table}
\section{The confrontation with the present-day orbit accuracy}
We will focus our attention to the gravitoelectric post-Newtonian
periselenium advance which is more than 3 orders of magnitude
larger than the gravitomagnetic effects, as can be inferred from
Tab. 3. Indeed, it should be noticed that the present-day quality
of force models applicable to low lunar satellites does not allow
highly precise determination of their orbits. The largely unknown
gravitational potential of the Moon remains the most important
perturbation force acting upon low lunar satellites (Floberghagen
et al., 1999).

The relativistic gravitoelectric perturbation affects only the
pericenter of the orbit of a test particle, as can be inferred
from a straightforward calculation based on the radial, transverse
and cross-track components of the related disturbing acceleration
(Grafarend and Joos, 1992). This implies that the relativistic
effect of interest can be mapped onto an along-track shift.
Indeed, in general (Christodoulidis et al., 1988) the along-track
perturbation can be written as \eqi \Delta s =
a\sqrt{1+\frac{e^{2}}{2}}\left[\Delta{\mathcal{M}}+\Delta\omega +
\Delta\Omega \cos i+\sqrt{(\Delta
e)^{2}+(e\Delta{\mathcal{M}})^{2}}\right].\eqf The element
${\mathcal{M}}$ is the mean anomaly. In the case of the
gravitoelectric pericenter shift it reduces to \eqi \Delta s_{\rm
GE} = a\sqrt{1+\frac{e^{2}}{2}}\Delta\omega_{\rm GE}.\eqf For Main
Orbiter and Rstar it yields the effects quoted in Tab. 4 for
different arc lengths which are of common use, e.g., in typical
lunar gravity field mapping missions.
\begin{table}[h]
\caption{Along-track relativistic gravitoelectric effects on the
SELENE sub-satellites.}
\begin{center}
\begin{tabular}{llll}
\noalign{\hrule height 1.5pt} Arc length & Main Orbiter &
Rstar\\
 \hline $1$ week& $8.6\times 10^{-2}$ m& $5.6\times 10^{-2}$ m\\
$2$ weeks& $1.7\times 10^{-1}$ m& $1.1\times 10^{-1}$ m\\
$4$  weeks& $3.7\times 10^{-1}$ m& $2.4\times 10^{-1}$ m\\
\noalign{\hrule height 1.5pt}
\end{tabular}
\end{center}
\end{table}
In order to investigate if such effects are within the present-day
or near forthcoming orbit accuracy of lunar missions we rely on
the results presented in (Floberghagen et al., 1999). In it the
RMS of orbit fit for a test case where the satellite polar orbits
are determined through low-low satellite-to-satellite tracking
(SST) in combination with Earth-based ground tracking is
presented. As input models for the gravity field and the albedo
the GLGM-2 lunar gravity model (Lemoine et al., 1997) and DLAM-1
lunar albedo model (Floberghagen et al., 1999), respectively, are
adopted.  It should be noticed that, due to the non-conservative
nature of the albedo force, the longer the arc, the larger the
orbit error. For the integrated SST Doppler measurements a
sampling rate of 20 s is assumed with a precision of $10^{-1}$ mm
s$^{-1}$. For arc lengths of one week of a polar 100 km altitude
satellite the along-track RMS is of the order of some meters or
even of $10^{1}$ m, as can be inferred from Tab. 2 of
(Floberghagen et al., 1999). Unfortunately, it is almost 1-2
orders of magnitude larger than the general relativistic
gravitoelectric effect on SELENE Main Orbiter quoted in Tab. 4.

These results seem to be confirmed also by the estimates of the
contribution of differential VLBI ($\Delta$VLBI) to the standard
2-way Doppler which allows for three dimensional orbit
determination of the SELENE Main Orbiter summarized in Fig. 2 of
(Heki et al., 1999). Over an arc length of 12 hours the
along-track accuracy amounts almost to 1 m, while the along-track
gravitoelectric relativistic shift, over the same time span is
$6.2\times 10^{-3}$ m.

In order to get an order of magnitude of the main systematic error
which could affect such kind of measurement we will calculate the
mismodelled classical precession of the periselenium of Main
Orbiter induced by the first even zonal harmonic $J_2$ of the
Moon's gravity field according to the LP75G lunar gravity model
(Konopoliv et al., 1998). The rate of the classical pericenter
advance due to the first even zonal harmonic is
\eqi\dot\omega_{\rm class}=-\frac{3n
J_2}{4(1-e^{2})^{2}}\left(\frac{R}{a}\right)^{2}\left(1-5\cos^{2}i\right).\eqf
According to Tab. 1 and Tab. 2 we have $\delta\dot\omega_{\rm
class}=5.1\times 10^{-11}$ s$^{-1}$ which is 3 orders of magnitude
larger than the relativistic gravitoelectric periselenium shift,
as can be inferred from Tab. 3. Anyway, this specific kind of
problems could be overcome by adopting suitable orbital residuals
combinations, as done in the terrestrial field for the
Lense-Thirring LAGEOS experiment (Ciufolini, 2000). However, it
should be also considered that, in the case of an orbital motion
around the Moon the long--term perturbations due to the odd zonal
harmonics of the lunar gravitational field play a role much more
important that in the Earth case (Kne${\rm \check{z}}$evi${\rm
\acute{c}}$ and Milani, 1998), especially for the eccentricity $e$
and the pericenter $\omega$.
\section{Conclusions}
The present-day level of accuracy of satellite selenodesy does not
allow the detection of the general relativistic gravitomagnetic
and gravitoelectric precessions of the satellites' orbits in the
field of the Moon. For example, such effects on the SELENE
subsatellites are, at most, 1 or 2 orders of magnitudes smaller
than the obtainable orbit accuracy. This holds for the
gravitoelectric periselenium advance of Main Orbiter. The
systematic errors which would be induced by the poor knowledge of
the lunar gravity field are almost 3 orders of magnitude larger
than the gravitoeletric periselenium advance of Main Orbiter.
%
\acknowledgments{ I wish to thank J. Ping for the useful
information kindly given me and A. Sengoku for the attention and
encouragement to this work.}

\begin{thebibliography}{99}
\bibitem{}
Christodoulidis, D., D. E. Smith, R. G. Williamson and S. M.
Klosko (1988):
 Observed tidal braking in the
 Earth/Moon/Sun system, J. Geophys. Res., (B6), 6216-6236.
\bibitem{}
Ciufolini, I. and J. A. Wheeler (1995): Gravitation and Inertia,
Princeton University Press, 498p.
\bibitem{}
Ciufolini, I. (2000): The 1995-99 measurements of the
Lense-Thirring effect using laser-ranged satellites, Class.
Quantum Grav., \textbf{17}(12), 2369-2380.
\bibitem{}
Ciufolini, I. (2001): Lense--Thirring effect and de Sitter
precession, in {Pascual-S${\rm \acute{a}}$nchez}, J.F., L.
{Flor${\rm\acute{\i}}$a}, A. {San Miguel} and F. {Vicente} (eds),
{Proceedings of the XXIII Spanish Relativity Meeting on Reference
Frames and Gravitomagnetism}, World Scientific, Singapore, 367p.
\bibitem{} Everitt, C.W.F., and other members of the Gravity Probe
B team, Gravity Probe B (2001): Countdown to Launch, in
{L$\ddot{\rm a}$mmerzahl}, C., C.W.F. {Everitt} and F.W. {Hehl}
(eds), Gyros, Clocks, Interferometers...:Testing Relativistic
Gravity in Space, Lecture Note in Physics 562, Springer Verlag,
Berlin, 507p.
\bibitem{} Floberghagen R., P. Visser, and F. Weischede (1999): Lunar
albedo force modeling and its effect on low lunar orbit and
gravity field determination, Adv. Sp. Res. \textbf{23}(4),
733-738.
\bibitem{} Grafarend, E. W., and G. Joos (1992):
Relativistic computation of geodetic satellite orbits, in
Linkwitz, K., and U. Hangleiter (eds), Proceedings of the 21nd
International Workshop on High Precision Navigation, 19-29, D${\rm
\ddot{u}}$mmler Verlag, Bonn.
\bibitem{} Heki, K., K. Matsumoto, and R. Floberghagen (1999):
Three-dimensional tracking of a lunar satellite with differential
very-long-baseline-interferometry, Adv. Sp. Res. \textbf{23}(11),
1821-1824.
\bibitem{} Kne${\rm \check{z}}$evi${\rm \acute{c}}$, Z., and A.
Milani (1998): Orbit maintenance of a lunar polar orbiter, Planet.
Space Sci., {\bf 46}(11/12), 1605--1611.
\bibitem{} Konopoliv, A.S., A.B. Binder, L.L. Hood, A.B.
Kucinskas, W.L. Sjogren, and J.G. Williams (1998): Improved
Gravity Field of the Moon from Lunar Prospector, Science,
\textbf{281}, 1476-1480.
\bibitem{} Konopoliv, A.S., S.W. Asmar, E. Carranza, W.L. Sjorgen
and D.N. Yuan (2001): Recent Gravity Models as a Result of the
Lunar Prospector Mission, Icarus, {\bf 150}, 1--18.
\bibitem{} Lemoine, F.G., M. Zuber, G. Neumann, and D. Rowlands
(1997): A 70$^{\rm th}$ degree lunar gravity model (GLGM-2) from
Clementine and other tracking data, J. Geophys. Res. (Planets)
\textbf{102}(E7), 16339-16359.
\bibitem{}
Lense, J. and  H. Thirring (1918): \"{U}ber den Einfluss der
Eigenrotation der Zentralk{\"{o}}rper auf die
 Bewegung der Planeten und Monde nach der Einsteinschen
Gravitationstheorie, Phys. Z., \textbf{19}, 156-163,  translated
by Mashhoon, B., F. W. Hehl and D. S. Theiss (1984): On the
Gravitational Effects of Rotating Masses: The Thirring-Lense
Papers, Gen. Rel. Grav., \textbf{16}, 711-750.
\bibitem{} Mashhoon, B., and D.S. Theiss (1991): Relativistic Lunar
Theory, Il Nuovo Cimento {\bf 106B}, 545-571.
\bibitem{} Mashhoon, B., and D.S. Theiss (2001): Relativistic
Effects in the Motion of the Moon, in {L$\ddot{\rm a}$mmerzahl},
C., C.W.F. {Everitt} and F.W. {Hehl} (eds), Gyros, Clocks,
Interferometers...:Testing Relativistic Gravity in Space, Lecture
Note in Physics 562, Springer Verlag, Berlin, 507p.
\bibitem{} Nordvedt, K (2001): Lunar Laser Ranging,
in {L$\ddot{\rm a}$mmerzahl}, C., C.W.F. {Everitt} and F.W. {Hehl}
(eds), Gyros, Clocks, Interferometers...:Testing Relativistic
Gravity in Space, Lecture Note in Physics 562, Springer Verlag,
Berlin, 507p.
\bibitem{} Ohta, K., K. Yonekura, Y. Takizawa, R. Nagashima, Y.
Iijima and S. Sasaki (2000): System Concept and Status of the
Selenological and Engineering Explorer (SELENE), Proc. 51st Int.
Astronautical Congress, IAF-00-Q.4.04.
\bibitem{}
Shapiro, I. (1990): Solar system tests of general relativity:
recent results and present plans, in Proceedings of the 12th
International Conference on General Relativity and Gravitation,
University of Colorado at Boulder, 1989, edited by N. Ashby, D.
Bartlett, and W. Wyss, Cambridge University Press, Cambridge,
313-330.

\end{thebibliography}
\end{document}